# Spin Filtering through Single-Wall Carbon Nanotubes Functionalized with Single-Stranded DNA

K. M. Alam and S. Pramanik
Department of Electrical and Computer Engineering
University of Alberta, Edmonton, AB T6G 2V4, Canada

## Abstract.

High spin polarization materials or spin filters are key components in spintronics, a niche subfield of electronics where carrier spins play a functional role. Carrier transmission through these materials is "spin selective" i.e. these materials are able to discriminate between "up" and "down" spins. Common spin filters include transition metal ferromagnets and their alloys, with typical spin selectivity (or, polarization) ~ 50% or less. Here we consider carrier transport in an archetypical one-dimensional molecular hybrid in which a single wall carbon nanotube (SWCNT) is wrapped around by single stranded deoxyribonucleic acid (ssDNA). By magnetoresistance measurements we show that this system can act as a spin filter with maximum spin polarization approaching ~ 74% at low temperatures, significantly larger than transition metals under comparable conditions. Inversion asymmetric helicoidal potential of the charged ssDNA backbone induces a Rashba spin-orbit interaction in the SWCNT channel and polarizes carrier spins. Our results are consistent with recent theoretical work that predicted spin dependent conductance in ssDNA-SWCNT hybrid. Ability to generate highly spin polarized carriers using molecular functionalization can lead to magnet-less and contact-less spintronic devices in the future. This can eliminate the conductivity mismatch problem and open new directions for research in organic spintronics.

# 1. Introduction.

Spin filters are solid-state systems that permit spin selective transmission of charge carriers and hence such systems are often used to generate and detect non-equilibrium spin populations in non-magnetic solids[1,2]. In the field of spintronics the central idea is to use spin as a state variable for information processing and storage, and hence spin filters are indispensable components in all-electrical spintronic devices[3,4]. Commonly used spin filter materials include transition metal ferromagnets (Ni, Fe, Co etc.) and their alloys (NiFe, CoFeB etc.)[1,2]. These materials are characterized by unequal density of states (DOS) of majority and minority spins, say $\rho_M$ and $\rho_m$ respectively, at the Fermi level (FL, Figure 1) and spin polarization ($P$) of these materials is often defined as the normalized difference of the spin up and spin down DOS[1]. Spin filtering through these materials can be understood by considering a simple model consisting of a ferromagnet/tunnel barrier junction as shown in Figure 1[1]. For generation (or injection) of spin-polarized carriers, consider the case when a negative bias is applied on the ferromagnet. In this case, electrons are injected from the ferromagnet through the tunnel barrier (indicated by right arrow in Figure1, top panel). Since the carriers at the (quasi) Fermi level are primarily responsible for tunneling and since these carriers have definite spin-polarization (parallel and antiparallel to the magnetization of the ferromagnet), the tunneling current injected from the ferromagnet will consist of carriers with definite spin polarizations. The population of carriers with spins parallel to majority (minority) spin should be proportional to $\rho_M$ ($\rho_m$) and hence the net injected spin polarization will be proportional to $\rho_M - \rho_m$ (assuming same tunneling probability for spin up and spin down states[1]). For diamagnetic materials (such as Au, Ag etc.), $\rho_M = \rho_m$ and hence these materials cannot be used to generate (or, inject) spin-polarized carrier population.

Similarly, a ferromagnet can be used to detect an incoming population (left arrow in Figure 1) of spin-polarized carriers. Suppose $n_M$ and $n_m$ are the densities of the incoming carriers with spins parallel to the majority and minority spins respectively, for a given polarization of the ferromagnet (Figure 1). In this case, tunnel conductance will be proportional to $n_M \rho_M + n_m \rho_m$[1]. However, if the magnetization of the ferromagnet is flipped, tunnel conductance will be proportional to $n_M\rho_m + n_m\rho_M$ [1]. Thus the incoming carriers will experience different resistances depending on the magnetization of the ferromagnet. Clearly, if the incoming carriers have no net spin polarization (i.e. $n_M = n_m$) then no change in device resistance will occur. Similarly, if we use a nonmagnet ($\rho_M = \rho_m$) instead of a ferromagnet, then also no change in device resistance will occur and no spin detection will take place. Thus, observation of a change in device resistance that depends on the magnetization orientation of the ferromagnetic electrode implies detection of spin-polarized carriers. Of course, it is important to ensure via control experiments that any such change is not due to other artifacts in the system.

An ideal spin filter, by definition, transmits only one particular spin polarization and blocks all others resulting in 100% spin selectivity. Such ideal spin filters are desirable since they can significantly improve the performance of technologically promising spintronic devices including magnetic tunnel junctions (MTJs)[1] and spin-field effect transistors (spin-FETs)[4]. For example, ideal spin filters will result in large magnetoresistance in MTJs, which is necessary for high density spintronic memories[1]. For spin-FETs, ideal spin filters can drastically improve the conductance on-off ratio by suppressing the leakage current[4]. Unfortunately, spin filtering using transition metal ferromagnets is far from ideal. For example, spin polarization of electrons tunneling through an alumina barrier is ~33% (nickel), 42% (cobalt), 45% (iron) etc.[1] Half metals have been proposed as an alternative, but unfortunately spin filtering efficiency of half metals is extremely sensitive to temperature and inhomogeneity at surface, interface and bulk [5]. Ferromagnetic tunnel barriers present another emerging platform for realization of spin filters[6].

Thus improvement of spin filtering efficiency remains an open problem in spintronics. Very recently several non-conventional materials have been investigated as potential spin filters. For example, it has been proposed theoretically that few layers of graphene as grown on Ni, can act as a "nearly perfect" spin filter[7]. This prediction is based on energy band alignment between graphene and Ni at the Fermi level, which allows transmission of primarily minority spins. However, to date this effect has not been demonstrated experimentally. Spin selective properties of chiral double-stranded (ds) DNA molecules have been investigated experimentally both by optical and electrical techniques[8,9]. In ref.[8] dsDNA monolayer was adsorbed on (non-magnetic) gold (Au) substrate. Spin polarization of the photoelectrons transmitted through the dsDNA layer was found to be ~ 60%[8]. In ref.[9] current flow has been measured in a nickel/dsDNA/Au device. A change in device resistance has been observed due to spin filtering through the dsDNA layer[9]. The physical process underlying the above-mentioned effect has been explored by several groups[10–13]. It has been suggested[10–13] that the inversion asymmetry of the helical DNA molecule can potentially enhance the effective spin-orbit interaction, despite the fact that atomic spin-orbit interactions are weak in DNA molecule. Using the spin-filtering property of chiral molecules, ref.[14] has recently demonstrated spin injection in a magnetic tunnel junction (MTJ) device. Here, one of the ferromagnets was replaced by [Au/chiral molecule] composite, which acted as a source of spin-polarized carriers. Spin-filtering property of chiral molecules has also been used to create optically induced local magnetization[15].

Another candidate that has been proposed theoretically as a spin filter is single wall carbon nanotube (SWCNT), wrapped by a chiral molecule (such as single-stranded (ss)-DNA), which creates a helical potential in the tube[16,17]. Carbon nanotubes (CNTs), which can be loosely described as single graphitic sheet rolled into a cylinder, have been explored extensively in the area of spintronics[18]. In almost all studies, carbon nanotubes have been used as a transport channel for spin polarized carriers and significant effort has been invested in understanding spin relaxation processes in these structures[18]. Since the nanotubes are made solely by low atomic number (carbon) atoms, spin-orbit coupling is expected to be extremely weak in these systems. However, recent studies have indicated that the curvature arising from the cylindrical geometry can enhance spin-orbit coupling due to hybridization among neighboring carbon atoms[19,20]. Further, the strength of the spin-orbit interaction is strongly dependent on the substrate and the species adsorbed on the graphitic surface[21].

Single stranded DNA (ssDNA) is known to interact strongly with carbon nanotubes and typically result in a helical wrapping of the nanotubes due to π-stacking (between DNA bases and the carbon atoms) and electrostatic interaction with charged sugar-phosphate backbone[22,23]. The ssDNA strands wrapped around the nanotube create a strong and inversion asymmetric electric field in the tube channel. For the charge carriers in the tube, this electric field transforms into a strong Rashba-type spin-orbit interaction. According to a recent theoretical work[16], such interactions can polarize the spins transmitted by ssDNA wrapped carbon nanotubes. To the best of our knowledge, no spin filtering experiment has been performed on ssDNA-wrapped carbon nanotubes. This is exactly what we intend to explore in this work.

The ssDNA-SWCNT hybrid has the added advantage of higher conductivity compared to pure DNA-based devices[8,9]. The DNA molecule is essentially an insulator, which makes it unsuitable for integration with practical spintronic devices, where the actual transport channel will most likely have smaller resistance. Further ssDNA-SWCNT hybrids allow efficient fabrication of planar spintronic devices, which is difficult to achieve using DNA molecules alone. The vertical geometry that is generally achieved by DNA molecules[8,9], is prone to pinhole shorts and possible creation of defects while depositing the top metallic contact. Such problems do not arise in the planar geometry.

In this work we have used the same principle described in the introductory paragraphs to examine the spin state of the carrier ensemble emanating from one end of ssDNA wrapped SWCNT. The other end of the ssDNA wrapped SWCNT is connected to a (diamagnetic) Au electrode (Figure 2a). The carriers are injected from the Au electrode to the SWCNT, which is wrapped by ssDNA. The carriers injected from the (diamagnetic) Au electrode have no net spin polarization. These carriers traverse through ssDNA wrapped SWCNT and a nickel electrode detects their net spin-polarization at the other end of the tube. It has been found that device resistance depends on the magnetization orientation of the Ni spin detector. Combined with various control experiments we show that this phenomenon is related to spin polarization of the charge carriers emanating from the ssDNA wrapped SWCNT. This work shows that ssDNA wrapped SWCNT can perform as a spin filter in the sense that it can polarize an injected population of spin-unpolarized carriers (which, in this case, originate from Au electrode). We further show that ssDNA wrapped SWCNTs are also able to detect the net spin polarizations of the carriers originating from a known spin injector (which, in this case, is Ni).

**2. Device Synthesis.**

We have fabricated ssDNA-SWCNT hybrids using a technique modified from the literature[22,23]. Details of our fabrication process have been described in the Supplementary Information (SI). Briefly, we start out with commercially available (Unidym Inc.) pure grade, semiconducting-dominant, bundled HiPco (high-pressure carbon monoxide) SWCNTs with nominal diameter of ~1 nm. Detailed characterizations of these tubes have been described in section 1 of SI. These bundled tubes are sonicated and dispersed in a nuclease-free aqueous solution (0.8mg SWCNT/1ml solution). This solution contains IDTE buffer (Tris and EDTA, pH 7.5, Integrated DNA Technologies Inc.) and ssDNA d(GT)$_{15}$ (1mg/1ml solution). According to molecular modeling studies[22,23], inter-tube binding energy is smaller than the binding energy between tube and ssDNA, which results in separation of the bundled tubes. Binding between helical ssDNA strands and nanotube takes place via π-stacking between the ssDNA bases and the $p_z$ orbitals of the carbon atoms on the tube surface as well as electrostatic interaction with charged sugar-phosphate backbone, which result in a stable helical wrapping of the tube by ssDNA[22,23].

The ssDNA-SWCNT solution is centrifuged to remove large impurities. After centrifugation, a drop of the decanted supernatant solution (containing ssDNA wrapped tubes) is cast on photolithographically patterned metal electrodes (Ni, Au on SiO$_2$), separated by ~ 750 nm, followed by rinsing (DI water), nitrogen drying and high temperature vacuum annealing (200°C, 30 minutes). Proper choices of ssDNA solution concentration, sonication parameters, and electrode separation ensure that the electrodes are bridged by few nanotubes. This is further confirmed by FESEM images (Figure 2(b), and SI). The thorough rinsing step removes any excess ssDNA and buffer molecule (Tris, EDTA) and thermal annealing step ensures reliable electrical contacts with reproducible characteristics and also removes water molecules, which can significantly affect transport behavior[24]. As described in the SI (section 2), thermal annealing also results in "tighter" wrapping of the tubes by ssDNA. We note that while ssDNA-SWCNT hybrids are unstable above 80°C in aqueous solutions, the critical temperature for these hybrids adsorbed on $SiO_2$ is much higher[25]. Using a binding energy value of ~ 0.7eV for ssDNA-SWCNT hybrids[25], the critical temperature for instability can be estimated to be ~ few 1000K, which is significantly larger than our annealing temperature. Device schematic and FESEM image of a working device are shown in Figures 2(a) and (b) respectively. Additional FESEM images are available in the SI, section 2.

Dispersion of SWCNTs in *excess* ssDNA solution (i.e. 0.8 mg SWCNT per 1mg ssDNA) as mentioned above, allows complete isolation of the bundled tubes and ensures ssDNA wrapping along the entire length of the nanotubes. This has been confirmed by the periodic height modulation pattern on the surface of the nanotubes observed in AFM (SI, section 2). Each turn of the ssDNA strand around the tube generates a peak in the AFM height measurement, resulting in a periodic height modulation pattern along the length of the nanotube. From the AFM images (SI), the average height of the nanotube top surface (measured from the substrate) is estimated to be ~ 2 nm. This is consistent with the average nanotube diameter (~ 1nm), helically wrapped by ssDNA strand which is separated from the top and bottom surfaces of the tube by ~ 0.5 nm due to π-stacking[26]. Thus the AFM height data suggests that ssDNA strands are closely arranged end-to-end in a single layer along the entire length of nanotubes. This is consistent with previous studies on ssDNA-SWCNT hybrids[22,23,25,26].

Figures 2(c), (d) show the Raman spectrum of ssDNA-SWCNT hybrids taken at 532nm excitation. As expected, the hybrid nanotubes have upshifted (~ 2–5 cm$^{-1}$) radial breathing mode (RBM) frequencies compared to pristine tubes (Figure 2(c)), which is a typical signature of ssDNA wrapping and is consistent with literature[27–29]. Multi-peak feature of the RBM band is preserved, with minor changes in relative intensities. This indicates minor change in diameter distribution as a result of wrapping. The wrapping process makes nanotubes stronger and stiffer than unwrapped ones, which restricts Raman active radial breathing mode and results in an upshift in the RBM frequencies[29].

Figure 2(d) compares the G band feature of pristine (unwrapped) and hybrid (wrapped, annealed) nanotubes. The G band for both cases consists of two peaks: (a) the $G^+$ peak at higher frequency (~ 1590cm$^{-1}$) originates from the vibrations of carbon atoms along the tube axis and (b) the $G^-$ peak at lower frequency (~ 1570cm$^{-1}$) is associated with the circumferential vibrations[30]. The high frequency component is ubiquitous in carbon-based materials including single wall CNTs, multiwall CNTs, graphene and graphite, but the low frequency component is a distinct feature of SWCNTs, which arises due to curvature and confinement[30]. As shown in Figure 2(d), lineshape of the G band has narrowed significantly as a result of wrapping, which can be attributed to the loss of inter-tube interaction upon wrapping[29] and indicates isolation of the nanotubes from the bundle. Also from Figure 2(d) we note that the relative intensity of $G^-$ to $G^+$ band is reduced after DNA wrapping. The calculated relative intensity for pristine tubes is 0.67; while for the hybrid tubes this ratio is 0.42. This intensity reduction is a common feature of the ssDNA-SWCNT hybrid and implies suppression of vibrations of carbon atoms along the circumferential direction as a result of wrapping[27]. From Figure 2(d) we also notice that the $G^+$ peak of hybrid nanotubes has upshifted by ~ 10 cm$^{-1}$ as a result of wrapping. This feature is consistent with previous literature and is associated with charge transfer from nanotubes to DNA molecule[27,30], which further confirms tight physical wrapping of the tubes by ssDNA. As shown in SI, the G band does not show any signature of Breit-Wigner-Fano (BWF) lineshape even after wrapping, indicating negligible presence of metallic nanotubes. The current-voltage characteristics (reported in section 3) have shown semiconducting temperature dependence, consistent with the above picture. The Raman features of ssDNA-SWCNT hybrid confirm that there exists a strong interaction between ssDNA and SWCNT. Further details on device synthesis is available in SI, section 3.

## 3. Results and Discussion.

Detailed electrical measurements have been performed on the devices described above. Both two-point and four-point measurements have been employed. Difference between these two measurements is negligible (SI, section 4), indicating that the series resistance of the metallic contact pads does not play a major role in the transport measurements. Further, we have characterized individual thin film metal electrodes (not shown), which unlike actual devices show metallic temperature dependence in current-voltage characteristics. Also these electrodes have orders of magnitude lower resistance than the actual devices in the same bias range. Thus, we conclude that contact resistances play a negligible role in the overall device resistance.

There exist two possible conduction pathways between the metallic electrodes: the SWCNT and the ssDNA molecular chain that wraps the SWCNT. Leakage current via the $SiO_2$ substrate is below the detection threshold of our measurement setup. The buffer material (Tris, EDTA) consists of small molecular compounds and does not form bonds with ssDNA or SWCNT. The ssDNA used in this study has only 30 base pairs, and hence each strand is ~ 10 nm long. Therefore, neither Tris/EDTA nor any unused ssDNA are expected to bridge the metal electrodes, which have a nominal separation of 750 nm. This has been confirmed later by a control experiment. Extended rinsing ensures presence of only trace amounts of these materials (if any) between the electrodes.

In the past various groups have explored electrical transport in DNA strands with varying lengths[31–36]. In these molecules, two different conduction processes can take place[31–36]: (a) coherent tunneling between the electrodes and (b) diffusive thermal hopping between various molecular sites. In our devices the ssDNA molecular chain (consisting of many small strands) is long, ~ 750 nm, which precludes conduction via coherent tunneling. Hopping transport, assisted by phonons, is expected to be the dominant process for such long chain ssDNA molecules. However, this mechanism is ineffective at the low temperature range (~ 10–40K) employed in this study and under these operating conditions the long chain ssDNA molecule can be viewed as an insulator. This is consistent with the established literature on electrical conduction via long DNA chains[31–36]. The nitrogen drying step and extended (~ 30 minutes) high temperature (~200°C) annealing step described above make any water molecule mediated ionic conduction highly unlikely. Further, this mode of conduction will be even less effective at cryogenic temperatures and under dc operating conditions employed in this study. Thus we conclude that the SWCNT is the primary charge transport channel in our devices. Direct experimental evidence of this picture has been discussed later in this work, where we show that there is no qualitative difference between the temperature-dependent current-voltage characteristics of SWCNT and ssDNA-wrapped SWCNT. In fact, if the ssDNA chain participates in charge conduction, it results in a "staircase shape", with a significant conductance suppression (due to its insulating nature) over a bias range of ~ 1V near zero bias in the current-voltage characteristics[37]. Such effects have not been observed in our experiments even at room temperature (Figure 3(f)). Thus the ssDNA chain does not offer any direct charge transport pathway to the carriers in our experiments.

It is important to note the distinction between our devices and previous DNA-based spin filters as reported in ref.[9]. For example, in ref.[9] no SWCNT was present and spin selectivity of *double stranded* DNA molecule was explored. In this case charge transfer occurred via the double stranded DNA molecules. The current-voltage characteristics (taken at room temperature) in this work[9] indicates a wide bandgap semiconductor (or insulator) behavior, which is consistent with existent literature on charge transport in DNA[31]. In our case, as described above, the main conduction channel is SWCNT. The ssDNA molecular chain only offers a helicoidal potential to the charge carriers in the nanotube due to the electrostatically charged sugar-phosphate backbone. As discussed before, the helical wrapping of ssDNA around SWCNT is caused by the electrostatic interaction of the sugar-phosphate backbone, as well as overlap of π orbitals of ssDNA bases and ($sp^2$ hybridized) carbon atoms of the SWCNT (so called π stacking).

The current-voltage (*I-V*) measurements have been performed for in-plane magnetic field in the range ±1.2T, over a wide temperature range of 12–295K. Details of the measurement apparatuses are provided in SI. Typical measurement geometry is shown in Figure 2(a). Since the in-plane coercivities of the Ni thin film electrodes are very small (~ 90G, SI section 5), magnetization of the Ni electrode should get flipped while varying the field in the above-mentioned range. Device resistance (*R*) has been extracted from the measured *I-V* characteristics by numerical differentiation ($R = dV/dI$). As described below, various control devices have been fabricated and tested to obtain a better understanding of the observed behavior. Multiple (~10) devices have been tested in each category and the behavior reported below is representative of each category.

First, as shown in Figure 3(a)-(e), we have measured the current-voltage characteristics of Au/[ssDNA-SWCNT]/Ni devices for +1.2T and –1.2T in the temperature range of 12–40K. The key observation here is the manifestation of a splitting in the current-voltage characteristics as a function of the magnetic field orientation. This effect is strongest at the lowest measurement temperature of 12K and gradually decreases as the temperature is increased and almost vanishes at ~ 40K. The insets in Figure 3(a)-(e) show the numerically computed resistance (d$V$/d$I$) versus bias ($V$) at the corresponding temperatures. A clear gap in the resistance has been observed ("vertical split", for a given voltage) as a function of magnetic field orientation. The resistance values near zero voltage bias are not shown (insets) since the current values in this bias range are below the detection threshold of our measurement setup and hence are not reliable. Figure 3(f) shows the temperature dependence of the zero magnetic field current-voltage characteristics over the entire temperature range of 12–295K used in this study. Semiconducting temperature dependence has been observed, which is consistent with prior studies on SWCNTs[38].

To understand the evolution of device resistance with magnetic field, we have recorded *I-V* characteristics at 100 different field values in ± 1.2T range. As before, device resistance has been computed by numerical differentiation of these curves. Figure 4 shows device resistance ($R$) as a function of the external magnetic field ($B$) for two different values of voltage bias. A hysteretic magnetoresistance feature and a resistance switching behavior have been observed in this plot. For example, during the positive sweep (i.e. – 1.2T to 1.2T), resistance switching occurs at the positive quadrant around 90G (Figure 4, main images), but for the negative sweep, resistance switching occurs in the negative quadrant around –90G. The switching field of 90G corresponds well with the in-plane coercivity of Ni (SI, section 5). Amount of resistance change (Δ$R$) is significant, ~ 60% at 12K and 0.35V bias. Change in resistance decreases as bias voltage is increased. For example, at 12K and 1.5V, $\Delta R \sim 4\%$ as seen from Figure 4(b). Device resistance saturates at large magnetic field values (Figure 4, insets).

Clearly, as seen from Figures 3 and 4 a strong magnetoresistance effect exists in Au/[ssDNA-SWCNT]/Ni devices. This magnetoresistance can originate from various sources and in order to obtain a better understanding of this effect we have tested four types of reference devices. In the first type of reference device, the Ni electrode has been replaced by a gold electrode and everything else has been kept the same as before. Transport data from these reference devices are shown in SI, section 6 (Figure S6). In this case no splitting in the current-voltage characteristics has been observed even at the lowest temperature of 12K. Otherwise, these samples exhibit qualitatively similar current-voltage characteristics as in Figure 3. Thus we conclude that the presence of the Ni electrode is essential for observation of the splitting and the hysteretic resistance switching reported in Figures 3, 4. This reference experiment also confirms that there is no significant orbital magnetoresistance effect of Au and [ssDNA-SWCNT] and hence these effects are not responsible for the features observed in Figures 3, 4.

In the second control experiment (not shown), we have measured the current-voltage characteristics in a blank sample, in which drop casting has been performed just using Tris/EDTA/ssDNA (no SWCNT) solution followed by the same rinsing, drying and annealing sequence used for the actual samples. No measurable current has been detected, implying that any remnant Tris/EDTA molecules or unused ssDNA in our original device (Figure 3) do not play any major role in conduction.

In our third reference experiment we have explored the role of ssDNA wrapping. In this case Au/[SWCNT]/Ni devices have been fabricated in absence of ssDNA molecules and hence in absence of any wrapping. As purchased bundled tubes have been dispersed by a sequence of ultrasonication and centrifugation steps (similar to the hybrid tubes) in 1% sodium dodecyl sulfate[24] (instead of d(GT)$_{15}$ solution as in actual samples). As shown in Figure S6 (c), (d) (SI), the current-voltage characteristics are qualitatively in agreement with Figure 3. This provides direct experimental evidence in support of our previous observation, where we argued that the primary conduction channel in our devices is SWCNT and the ssDNA molecular chain has no direct role in conduction. The observed current-voltage characteristics in Figure S6(c), (d) is consistent with previous studies performed on SWCNTs of comparable dimensions and operating temperatures[38]. However, unlike Figure 3, we do not observe any splitting in the current-voltage characteristics when an external magnetic field is applied. Thus we conclude that wrapping of the SWCNT channel by ssDNA is a key component behind the observed splitting. No significant magnetoresistance effect has been observed in this case, implying negligible presence of any spurious Hall effect due to the fringing field lines from the Ni contact.

Since the Ni electrode appears to play an important role, in our final reference experiment we have characterized the magnetoresistance effect of the Ni electrodes (Figure S7). In this case we have observed a weak magnetoresistance effect (~ 1%, that arises due to the anisotropic magnetoresistance or AMR effect of Ni) and the device resistance values are roughly the same for +1.2T and – 1.2T. Further these electrodes exhibit linear current-voltage characteristics with metallic temperature dependence (not shown), which is in contrast with the features observed in Figure 3. Also, the resistance values of the Ni contacts are orders of magnitude smaller than the devices reported in Figures 3, 4. Thus we conclude that the Ni electrode resistance and corresponding magnetoresistance do not play a major role in the observed splitting.

In the past it has been proposed that certain chemical groups (such as thiols) when adsorbed on Au can make the Au surface magnetic[39,40]. Such effects are unlikely to occur in our experiments simply because we do not use thiol groups unlike the experiments cited above. The chemicals used in our experiment (Tris, EDTA) are not known to have similar properties, and the fabrication steps used in this study ensures presence of such chemicals by trace amounts only. Thus the features observed in Figures 3, 4 couldn't arise from these effects.

Summarizing the above discussion, we conclude that the splitting in the *I-V* characteristics is observed only when SWCNT is wrapped with ssDNA and one of the contacts is Ni. This effect has been shown not to originate from any orbital magnetoresistance effect of the tubes or the contacts or any spurious Hall effect in the system. Thus we proceed to explain this effect using the framework of spin-polarized carrier generation and detection. The model that we use is based on a recent theoretical work[16] that predicted spin polarized conductance for ssDNA-wrapped SWCNTs, which originates due to the Rashba spin-orbit coupling mediated by the inversion asymmetric helicoidal electric field of the ssDNA strands.

In our transport measurement setup, the nickel contact has always been used as the ground terminal. For negative voltage polarity on the Au electrode, electrons are injected into the ssDNA-SWCNT from the Au electrode across a Schottky-type potential barrier at the interface. Carrier injection occurs primarily via tunneling, at least in the low temperature regime reported in this work (12–40K) since thermionic emission is unlikely at such low temperatures. Since Au is diamagnetic, these electrons have no net spin polarization in any direction, even in the presence of an external magnetic field. In this ensemble of electrons, 50% of the population has "spin up" polarization component (using externally applied magnetic field as our reference direction) and remaining 50% has "spin down" polarization component and the net spin polarization is zero. As these electrons traverse through the ssDNA wrapped SWCNT, it has been shown in ref.[16] that they accumulate some net spin polarization (say *P*). If this happens, the accumulated spin polarization *P* can be sensed by the ferromagnetic Ni "spin detector" at the other end of the device. Note that there exists a Schottky-type barrier at this detecting interface as well. This is due to energy band misalignment at this interface as well as due to any native oxide of Ni. As the magnetization of the Ni electrode is flipped, as seen in Figures 3, 4, it indeed offers a different resistance to these carriers, indicating non-zero value of *P*. Thus, combining with the results from the reference experiments we conclude that ssDNA wrapped SWCNT indeed polarizes an initially spin-unpolarized population of carriers. It is important to note that this spin polarization is created by ssDNA wrapping and not by the applied magnetic field. This is also confirmed by the control experiments described below.

The reverse effect takes place when the voltage polarity is positive. In this case spin polarized electrons are injected from the Ni electrode. The Au/[ssDNA-SWCNT] composite can be viewed as an effective electrode with a fixed spin polarized density of states (with spin polarization $P$) at the (quasi) Fermi level. This composite now acts as a spin detector. As Ni magnetization is flipped, so does the polarity of the incoming spins from Ni. The composite electrode offers different resistances in both cases resulting in split current-voltage characteristics in the positive quadrant.

For Au/[ssDNA-SWCNT]/Au reference devices (Figure S6 (a), (b)), no spin detecting electrode is present since Au is diamagnetic. Therefore even though spin-polarized carriers are generated by the Au/[ssDNA-SWCNT] composite, they go undetected by the second Au electrode (as discussed in the introductory paragraphs) and hence no splitting is observed in the current-voltage characteristics. For Au/[SWCNT]/Ni devices (Figure S6 (c), (d)), no splitting has been observed, which underscores the critical role played by the helicoidal potential of the ssDNA strands. Just spin-orbit interaction of SWCNT alone or the applied magnetic field is not sufficient to generate a spin polarization.

Using the physical picture described above, we can estimate the spin polarization $P$ of the carriers emanating from ssDNA-SWCNT by applying Julliere formula[1]. As mentioned before, these spin polarized electrons are ultimately detected by the Ni electrode. There exists a Schottky-type potential barrier at the Ni/[ssDNA-SWCNT] interface, and at low temperatures employed in this study, transport at this interface is primarily due to tunneling, rather than thermionic emission. Applying Julliere formula at the tunnel junction at the detecting interface, and assuming 33% spin polarization for Ni[1], we obtain a spin polarization value (lower limit) of $P \sim 74\%$ at 12K and 0.35V, which decays gradually as the bias is increased (Figure 5a). The loss in spin polarization at increased bias can be explained by invoking spin relaxation. At higher bias, electrons are likely to scatter more during their transit through the SWCNT, which tends to randomize the accumulated polarization *P* in presence of spin-orbit interaction and cause spin relaxation. As a result *P* decreases with increasing bias. Near zero bias, current is negligibly small for both orientations of the magnetic field and is below the experimental detection level of our apparatus, and hence estimation of *P* is difficult in this bias range (Figure 5a). However, *P* shows sign of saturation near zero bias (Figure 5a), indicating that our measured *P* is close to the maximum value.

Loss of spin polarization at higher bias can also be explained using a simple model in which each spin has a different conduction barrier[9]. At very high bias both spins tunnel equally well, resulting in loss of net transmitted spin polarization. But at low bias only the low barrier spin will tunnel, resulting in higher transmitted spin polarization.

The disappearance of this effect at higher temperature (Figure 5b) can be ascribed to the enhanced spin relaxation within the SWCNT at higher temperatures. In order to observe the spin filtering effect the Rashba spin-split bands should remain sufficiently distinct. At higher temperature thermal energy broadens these bands, leading to enhanced spin mixing and loss of spin filtering. Variation of *P* as a function of temperature has been shown in Figure 5b.

From the computed d*V*/d*I* curves (Figure 3a-e insets) we observe that a given conductance level is achieved for different bias values, depending on the magnetization orientation of Ni. This bias differential (the "horizontal split" for a given conductance) should reflect the effective potential barrier experienced by the "unfavorable" spins as compared to the "favorable" ones. The barrier is ~ 0.2 eV for a bias of ~ 0.35V (and ~ 0.15eV for a bias of ~ 1.5V) and the barrier height decreases with bias (Figure 3a-e). Such spin-dependent barrier height is too high to be explained by invoking atomic spin-orbit coupling (~ few meV) of lightweight elements that comprise ssDNA-SWCNT hybrid. In the present case, SWCNT and ssDNA primarily consist of light elements with atomic spin-orbit coupling values at least two orders of magnitude smaller than the barrier height estimated above. Also as discussed previously, the main conduction channel is SWCNT and the carbon atoms, in their most abundant form ($^{12}C$), do not offer any hyperfine interaction. Thus these mechanisms alone cannot explain the large spin dependent barrier height discussed above. Therefore this work should be viewed within the framework of the Rashba spin-orbit interaction induced by the inversion asymmetric helicoidal electric field of the ssDNA[16,17]. It is to be noted that similar unusual large splitting between the two spin states has been observed in spin selective conduction through double stranded DNA[9].

**4. Conclusion.**

In conclusion, we have demonstrated spin-selective carrier transmission (or, spin filtering) using a non-magnetic system in which SWCNT is wrapped with ssDNA strands. Significant spin polarization, larger than typical transition metal ferromagnets, has been observed at low temperatures. The helical wrapping of the ssDNA induces an inversion asymmetric electric field in the channel, which results in a Rashba type spin-orbit interaction and polarizes carrier spins. The possibility of generating spin-polarized carriers via non-magnetic materials can result in "magnet-less" and "contact-less" spintronic devices in the future, which will alleviate the conductance mismatch problem that plagues almost all spintronic devices. Selective wrapping of a carbon nanotube network by ssDNA can result in localized sources of spin-polarized carriers, which may allow realization of spin-based circuits.

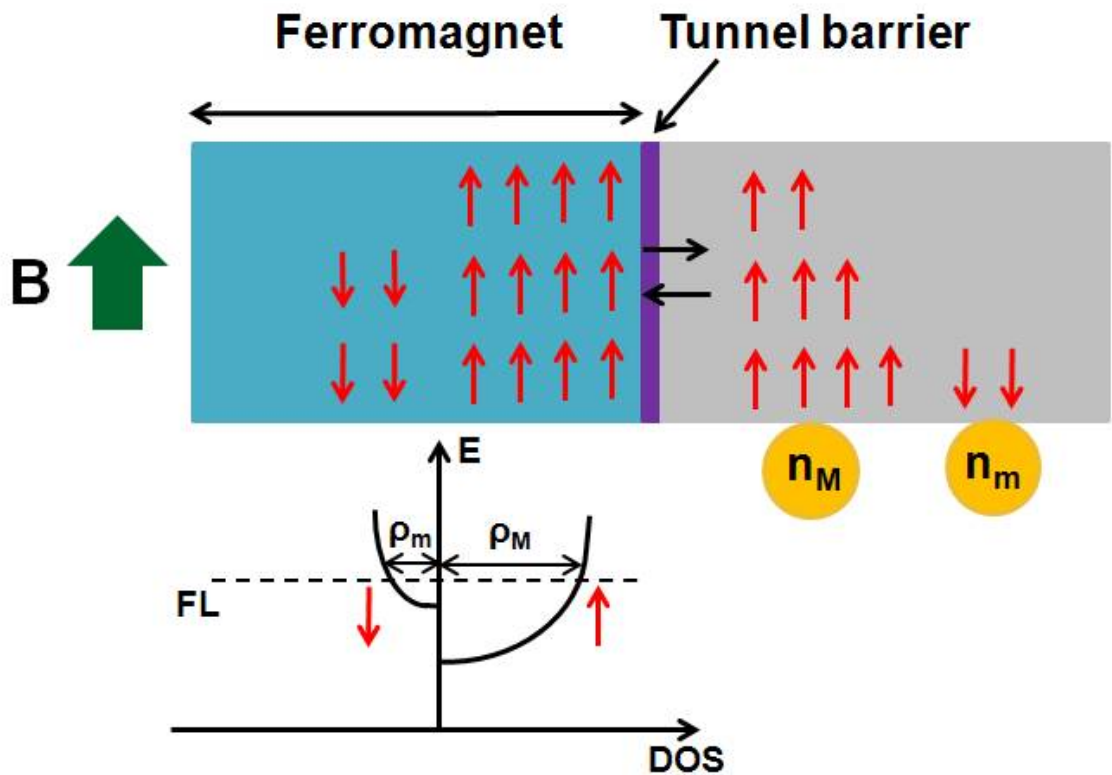

**Figure 1.** Schematic description of the spin injection and detection functions of a spin filter. In the top panel the right arrow indicates spin injection from the ferromagnet through the tunnel barrier. The left arrow indicates an incoming population of spin polarized carriers which will be detected by tunnel barrier/ferromagnet junction. Magnetization orientation of the ferromagnet is controlled by an external magnetic field **B**. The bottom panel represents energy ($E$) vs. density of states (DOS) of majority and minority spins of the ferromagnet. At the Fermi level (FL) these two DOS are different ($\rho_M \neq \rho_m$).

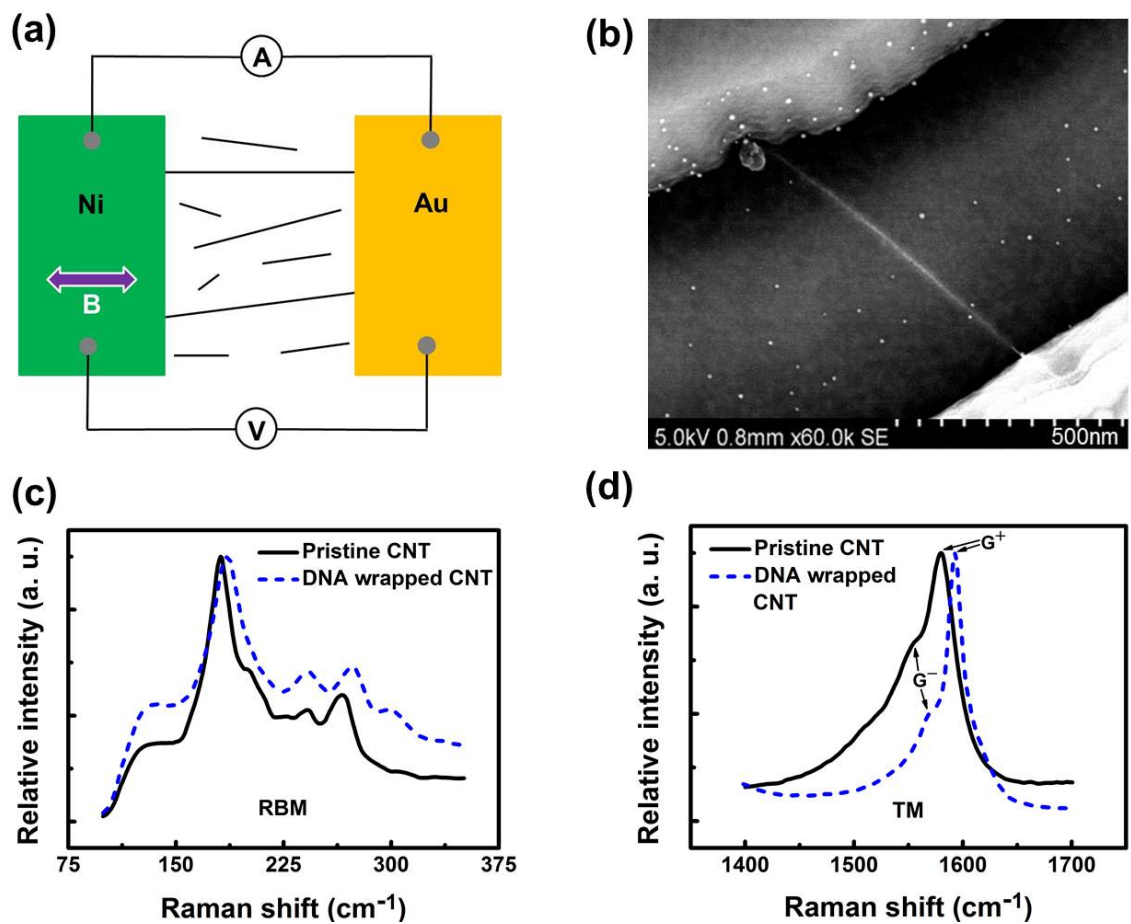

**Figure 2.** (a) Device schematic and four point measurement geometry. Very few ssDNA-SWCNTs are contacted by both electrodes. Magnetic field (***B***) is applied in plane of Ni thin film. (b) FESEM image of a working device showing a single ssDNA-SWCNT bridging the two contacts. (c) Raman RBM band (532 nm excitation) of as purchased (pristine) and hybrid (wrapped, annealed) tubes. (d) Raman G band (532 nm excitation) of as purchased (pristine) and hybrid (wrapped, annealed) tubes. As discussed in text, Raman data indicates wrapping of SWCNTs by ssDNA. Further proof of wrapping has been provided in SI.

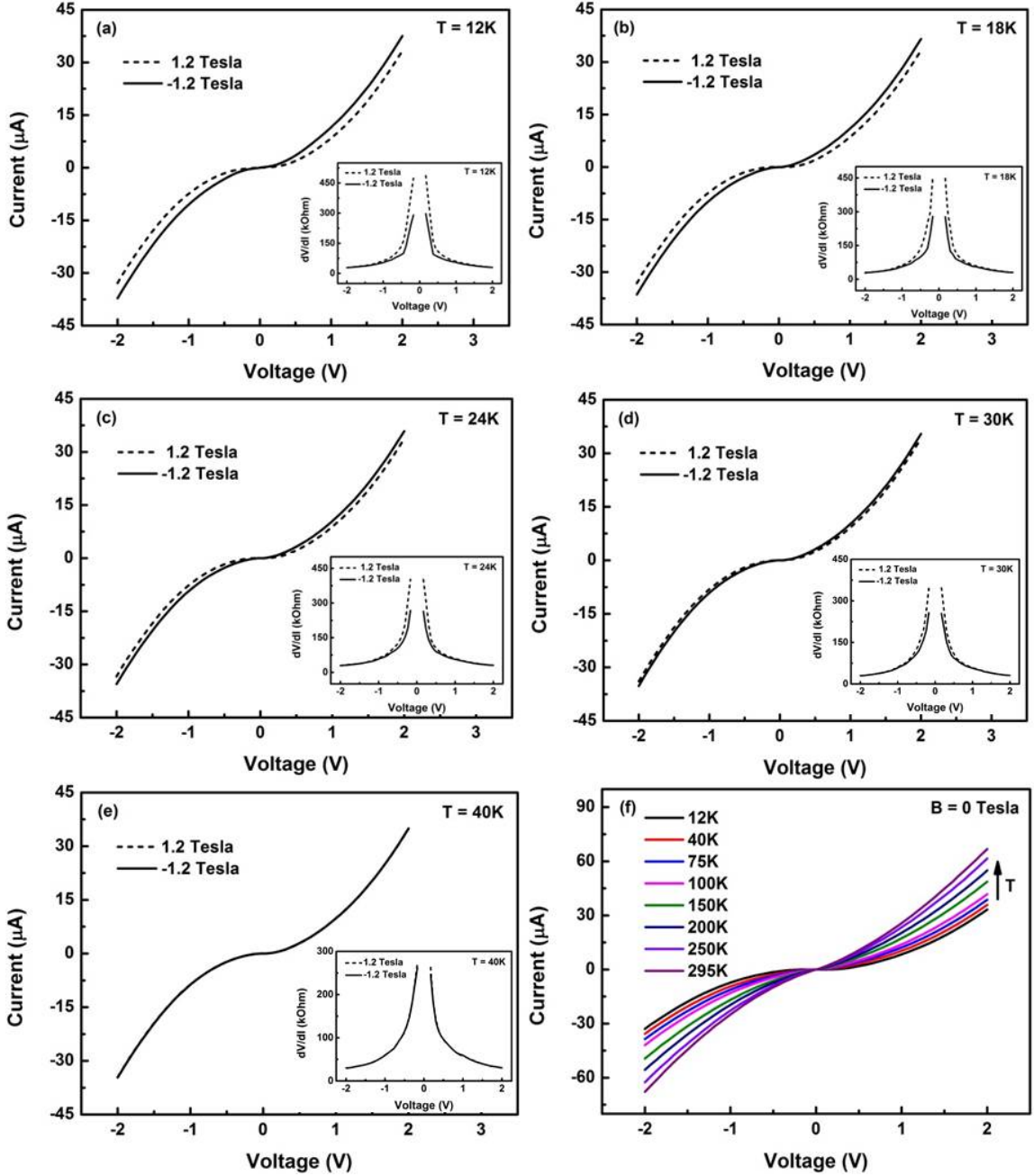

**Figure 3.** Temperature and magnetic field dependent current-voltage (*I-V*) characteristics of hybrid nanotubes connected between Ni and Au electrodes. The insets show magnetic field dependent d*V*/d*I* as a function of *V*. The low-bias (close to *V* = 0) d*V*/d*I* values are not plotted since the current values in this bias range are below the measurement threshold.

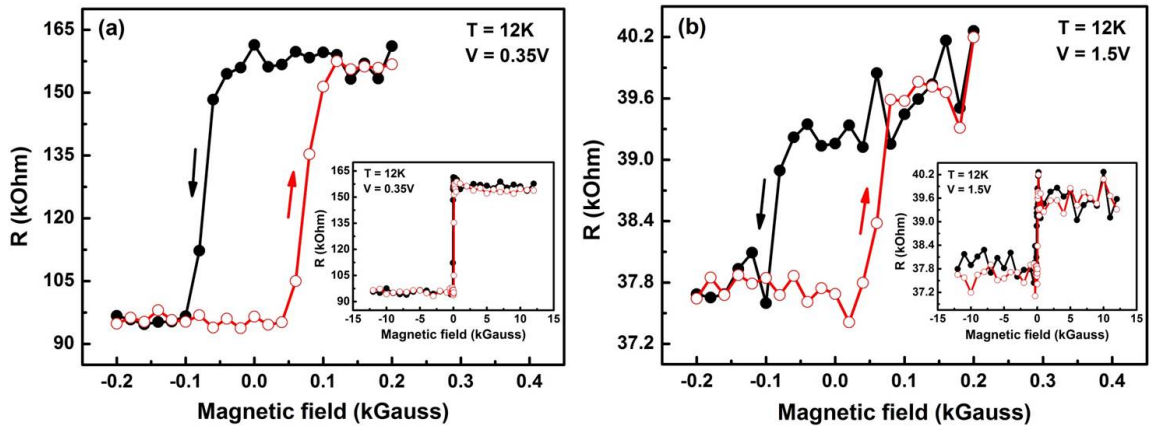

**Figure 4.** Resistance (*R*) vs. Magnetic field (*B*) plots at two different voltage biases. Hysteretic magnetoresistance feature has been observed, which becomes weaker at higher voltage bias. Switching magnetic field is ~ 90G.

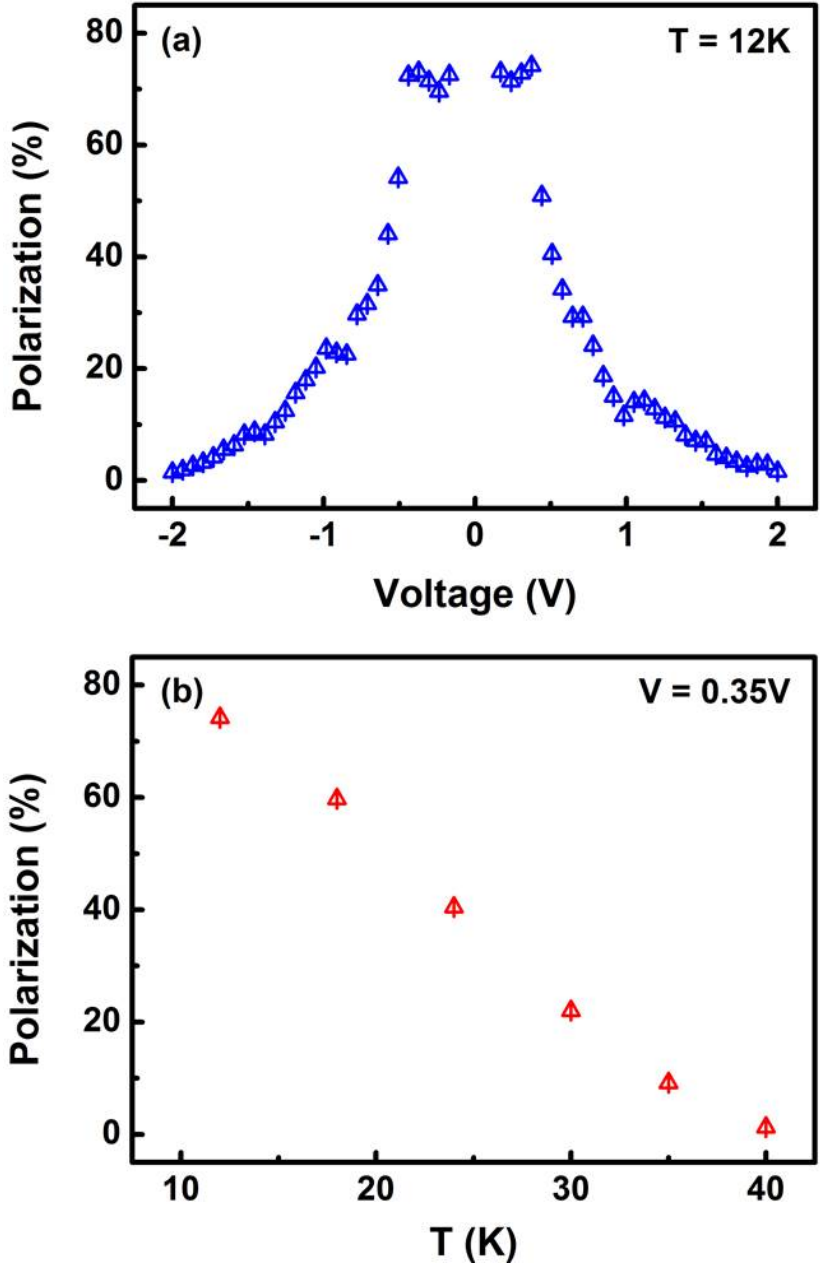

**Figure 5.** Computed polarization ($P$) as a function of bias ($V$) and temperature ($T$). (a) $P$ vs. $V$ at the lowest measurement temperature of 12K, at which the effect is strongest. Polarization decays with increasing bias and shows sign of saturation in the low bias limit ($\leq 0.35$ V). (b) $P$ vs. $T$ at bias of 0.35V, at which polarization is maximum (as seen from (a)). As seen from (b), polarization decays with increasing temperature.

[1]  E. Y. Tsymbal, O. N. Mryasov, P. R. LeClair, *J. Phys. Condens. Matter* **2003**, *15*, R109.
[2]  K. Inomata, N. Ikeda, N. Tezuka, R. Goto, S. Sugimoto, M. Wojcik, E. Jedryka, *Sci. Technol. Adv. Mater.* **2008**, *9*, 14101.
[3]  C. Chappert, A. Fert, F. N. Van Dau, *Nat. Mater.* **2007**, *6*, 813.
[4]  S. Bandyopadhyay, M. Cahay, *Nanotechnology* **2009**, *20*, 412001.
[5]  P. A. Dowben, R. Skomski, *J. Appl. Phys.* **2004**, *95*, 7453.
[6]  G.-X. Miao, J. Chang, B. A. Assaf, D. Heiman, J. S. Moodera, *Nat. Commun.* **2014**, *5*, DOI 10.1038/ncomms4682.
[7]  V. M. Karpan, G. Giovannetti, P. A. Khomyakov, M. Talanana, A. A. Starikov, M. Zwierzycki, J. van den Brink, G. Brocks, P. J. Kelly, *Phys. Rev. Lett.* **2007**, *99*, 176602.
[8]  B. Göhler, V. Hamelbeck, T. Z. Markus, M. Kettner, G. F. Hanne, Z. Vager, R. Naaman, H. Zacharias, *Science* **2011**, *331*, 894.
[9]  Z. Xie, T. Z. Markus, S. R. Cohen, Z. Vager, R. Gutierrez, R. Naaman, *Nano Lett.* **2011**, *11*, 4652.
[10]  S. Yeganeh, M. A. Ratner, E. Medina, V. Mujica, *J. Chem. Phys.* **2009**, *131*, 14707.
[11]  R. Gutierrez, E. Díaz, R. Naaman, G. Cuniberti, *Phys. Rev. B* **2012**, *85*, 81404.
[12]  A.-M. Guo, Q. Sun, *Phys. Rev. Lett.* **2012**, *108*, 218102.
[13]  A.-M. Guo, Q. Sun, *Phys. Rev. B* **2012**, *86*, 35424.
[14]  S. P. Mathew, P. C. Mondal, H. Moshe, Y. Mastai, R. Naaman, *Appl. Phys. Lett.* **2014**, *105*, 242408.
[15]  O. Ben Dor, N. Morali, S. Yochelis, L. T. Baczewski, Y. Paltiel, *Nano Lett.* **2014**, *14*, 6042.
[16]  G. S. Diniz, A. Latgé, S. E. Ulloa, *Phys. Rev. Lett.* **2012**, *108*, 126601.
[17]  A. A. Eremko, V. M. Loktev, *Phys. Rev. B* **2013**, *88*, 165409.
[18]  A. Cottet, T. Kontos, S. Sahoo, H. T. Man, M.-S. Choi, W. Belzig, C. Bruder, A. F. Morpurgo, C. Schönenberger, *Semicond. Sci. Technol.* **2006**, *21*, S78.
[19]  F. Kuemmeth, S. Ilani, D. C. Ralph, P. L. McEuen, *Nature* **2008**, *452*, 448.
[20]  G. A. Steele, F. Pei, E. A. Laird, J. M. Jol, H. B. Meerwaldt, L. P. Kouwenhoven, *Nat. Commun.* **2013**, *4*, 1573.
[21]  D. Ma, Z. Li, Z. Yang, *Carbon* **2012**, *50*, 297.
[22]  M. Zheng, A. Jagota, E. D. Semke, B. A. Diner, R. S. Mclean, S. R. Lustig, R. E. Richardson, N. G. Tassi, *Nat. Mater.* **2003**, *2*, 338.
[23]  M. Zheng, A. Jagota, M. S. Strano, A. P. Santos, P. Barone, S. G. Chou, B. A. Diner, M. S. Dresselhaus, R. S. Mclean, G. B. Onoa, G. G. Samsonidze, E. D. Semke, M. Usrey, D. J. Walls, *Science* **2003**, *302*, 1545.
[24]  M. Cha, S. Jung, M.-H. Cha, G. Kim, J. Ihm, J. Lee, *Nano Lett.* **2009**, *9*, 1345.
[25]  D. A. Yarotski, S. V. Kilina, A. A. Talin, S. Tretiak, O. V. Prezhdo, A. V. Balatsky, A. J. Taylor, *Nano Lett.* **2009**, *9*, 12.
[26]  J. F. Campbell, I. Tessmer, H. H. Thorp, D. A. Erie, *J. Am. Chem. Soc.* **2008**, *130*, 10648.
[27]  H. Kawamoto, T. Uchida, K. Kojima, M. Tachibana, *Chem. Phys. Lett.* **2006**, *432*, 172.
[28]  H. Kawamoto, T. Uchida, K. Kojima, M. Tachibana, *J. Appl. Phys.* **2006**, *99*, 94309.

[29] Q.-H. Yang, N. Gale, C. J. Oton, F. Li, A. Vaughan, R. Saito, I. S. Nandhakumar, Z.-Y. Tang, H.-M. Cheng, T. Brown, W. H. Loh, *Nanotechnology* **2007**, *18*, 405706.

[30] M. S. Dresselhaus, G. Dresselhaus, R. Saito, A. Jorio, *Phys. Rep.* **2005**, *409*, 47.

[31] D. Porath, A. Bezryadin, S. de Vries, C. Dekker, *Nature* **2000**, *403*, 635.

[32] A. J. Storm, J. van Noort, S. de Vries, C. Dekker, *Appl. Phys. Lett.* **2001**, *79*, 3881.

[33] C. Dekker, *Phys. World* **n.d.**, *August 2001*, 29.

[34] R. G. Endres, D. L. Cox, R. R. P. Singh, *Rev. Mod. Phys.* **2004**, *76*, 195.

[35] M. DiVentra, M. Zwolak, in *DNA Electron. Encycl. Nanosci. Nanotechnol.*, **2005**, pp. 475–493.

[36] J. R. Heath, M. A. Ratner, *Phys. Today* **2003**, *56*, 43.

[37] J. S. Hwang, H. T. Kim, H. K. Kim, M. H. Son, J. H. Oh, S. W. Hwang, D. Ahn, *J. Phys. Conf. Ser.* **2008**, *109*, 12015.

[38] M. Bockrath, D. H. Cobden, P. L. McEuen, N. G. Chopra, A. Zettl, A. Thess, R. E. Smalley, *Science* **1997**, *275*, 1922.

[39] P. Crespo, R. Litrán, T. C. Rojas, M. Multigner, J. M. de la Fuente, J. C. Sánchez-López, M. A. García, A. Hernando, S. Penadés, A. Fernández, *Phys. Rev. Lett.* **2004**, *93*, 87204.

[40] J. S. Garitaonandia, M. Insausti, E. Goikolea, M. Suzuki, J. D. Cashion, N. Kawamura, H. Ohsawa, I. Gil de Muro, K. Suzuki, F. Plazaola, T. Rojo, *Nano Lett.* **2008**, *8*, 661.